\documentclass{sigchi-ext}
\usepackage[T1]{fontenc}
\usepackage{textcomp}
\usepackage[scaled=.92]{helvet} 
\usepackage{graphicx} 
\usepackage{balance}  
\usepackage{booktabs} 
\usepackage{ccicons}  
\usepackage{ragged2e} 

\usepackage{cleveref}

\def\plaintitle{SIGCHI Extended Abstracts Sample File: Note Initial
  Caps} 
\def\emptyauthor{}
\def\plainkeywords{Authors' choice; of terms; separated; by
  semicolons; include commas, within terms only; required.}

\title{On the Effect of Information Asymmetry in Human-AI Teams
}

\numberofauthors{6}

\author{%
  \alignauthor{%
    \textbf{Patrick Hemmer*}\\
    \affaddr{Karlsruhe Institute of \\ Technology} \\
    \affaddr{Karlsruhe, Germany} \\
    \email{patrick.hemmer@kit.edu}}\alignauthor{%
    \textbf{Michael Vössing}\\
    \affaddr{Karlsruhe Institute of \\ Technology}\\
    \affaddr{Karlsruhe, Germany}\\
    \email{michael.voessing@kit.edu} } \vfil \alignauthor{%
    \textbf{Max Schemmer*}\\
    \affaddr{Karlsruhe Institute of \\ Technology} \\
    \affaddr{Karlsruhe, Germany} \\
    \email{max.schemmer@kit.edu}}\alignauthor{%
    \textbf{Gerhard Satzger}\\
    \affaddr{Karlsruhe Institute of \\ Technology}\\
    \affaddr{Karlsruhe, Germany}\\
    \email{gerhard.satzger@kit.edu} } \vfil \alignauthor{%
    \textbf{Niklas Kühl}\\    
    \affaddr{Karlsruhe Institute of \\ Technology}\\
    \affaddr{Karlsruhe, Germany}\\
    \email{niklas.kuehl@kit.edu} \\
    }
        \vfil
    \alignauthor{%
        * denotes equal contribution
    }
    }

\definecolor{linkColor}{RGB}{6,125,233}
\hypersetup{%
  pdftitle={\plaintitle},
  pdfauthor={\emptyauthor},
  pdfkeywords={\plainkeywords},
  bookmarksnumbered,
  pdfstartview={FitH},
  colorlinks,
  citecolor=black,
  filecolor=black,
  linkcolor=black,
  urlcolor=linkColor,
  breaklinks=true,
}

\begin{document}

\CopyrightYear{2022}
\setcopyright{rightsretained}
\conferenceinfo{CHI Conference on Human Factors in Computing Systems (CHI ’22), Workshop on Human-Centered Explainable AI (HCXAI)}{May 12--13, 2022, New Orleans, LA, USA}
\isbn{978-1-4503-6819-3/20/04}
\doi{https://doi.org/10.1145/3334480.XXXXXXX}
\copyrightinfo{\acmcopyright}

\maketitle

\RaggedRight{} 

\begin{abstract}
Over the last years, the rising capabilities of artificial intelligence (AI) have improved human decision-making in many application areas. Teaming between AI and humans may even lead to complementary team performance (CTP), i.e., a level of performance beyond the ones that can be reached by AI or humans individually. Many researchers have proposed using explainable AI (XAI) to enable humans to rely on AI advice appropriately and thereby reach CTP. However, CTP is rarely demonstrated in previous work as often the focus is on the design of explainability, while a fundamental prerequisite---the presence of complementarity potential between humans and AI---is often neglected. Therefore, we focus on the existence of this potential for effective human-AI decision-making. Specifically, we identify information asymmetry as an essential source of complementarity potential, as in many real-world situations, humans have access to different contextual information. By conducting an online experiment, we demonstrate that humans can use such contextual information to adjust the AI’s decision, finally resulting in CTP.
\end{abstract}



\section{Introduction}
The rising capabilities of artificial intelligence (AI) have paved the way for supporting human decision-making in a growing number of domains \cite{senoner2021using,treiss2020uncertainty,Wu2020DeepNN}. To offer humans meaningful support, particularly in high-stake settings, AI models are not only expected to provide accurate predictions but also a notion about how a decision was derived or how confidently it was made to foster humans' understanding. This idea fueled the development of techniques from the field of explainable AI (XAI) \cite{adadi2018}. Its intention is to enable domain experts to assess when to rely on AI advice to improve decision-making performance \cite{alufaisan2021does,bansal2021does,zhang2020effect}.

Ideally, this form of XAI-assisted decision-making achieves complementary team performance (CTP)---a task performance that surpasses both human and AI performance when conducting the task alone. However, current research reveals that achieving CTP is challenging \cite{bansal2021does,hemmer2021human}. Most studies show that XAI-assisted decision-making yields higher team performance than humans conducting the task alone. Still, this performance is often inferior to the one of the AI alone \cite{bansal2021does,hemmer2021human}, leaving the question unanswered why CTP could not have been accomplished. 

A possible explanation for this observation may be that in order for human-AI decision-making to result in CTP, a more fundamental prerequisite is the presence of sufficient complementarity potential (CP) between humans and AI. In this context, we hypothesize that a source of CP emerges from unique human contextual information (UHCI).
In practice, domain experts often have access to further information not available to the AI during training as not all data might be digitally available due to technical or economic reasons. Thus, we investigate whether humans' decision-making benefits from the presence of UHCI when receiving AI assistance. 

We conduct an online experiment within the domain of real estate appraisal. We employ an AI model that predicts real estate prices and provides an uncertainty estimate solely based on tabular data. Humans have additional access to a corresponding picture of the house and, thus, are equipped with UHCI. Our results demonstrate that the presence of UHCI can enable humans to adjust AI predictions resulting in a task performance that surpasses the one of humans and AI alone, i.e., CTP. From this finding we can derive several implications for future XAI research.

In general, sufficient CP might constitute a requirement for effective XAI-assisted decision-making. Therefore, researchers need to investigate the mutual effects between CP and XAI. On the one hand, CP may positively influence XAI-assisted decision-making. For example, UHCI may activate analytical instead of intuitive thinking and thereby could indirectly trigger conscious engagement with explanations which improves team performance.
On the other hand, XAI can also amplify the effect of CP. For example, feature importance can be used to detect whether the individual perceived UHCI is really unique or also taken into account by the AI.
In future work, we aim to formalize the notion of CP, evaluate the impact of different XAI types within our specific study setup, and further assess how humans can learn to rely on AI advice appropriately.

\section{Related Work}

In line with the continuous development of XAI algorithms \cite{adadi2018}, a growing body of research has started to investigate their effect on task performance in AI-assisted decision-making scenarios in online experiments. A popular idea is to enable humans to question the AI's decisions through insights about the uncertainty of the prediction \cite{bansal2021does,fugener2021will,zhang2020effect} or through explanations that aim to shed light on the AI's decision-making \cite{lai2020chicago,liu2021understanding}. In this context, studies analyze the effects of different XAI techniques, ranging from feature-based \cite{carton2020feature,mohseni2021machine,ribeiro2016should} over example-based \cite{nguyen2021effectiveness,van2021evaluating} to rule-based \cite{alufaisan2021does,ribeiro2018anchors} explanations, and also consider an entire spectrum between full human agency and full automation \cite{lai2019human}. Even though these studies reveal that human decision-making generally benefits from this algorithmic support, the combined human-AI performance usually remains inferior to the AI conducting the task alone, as humans struggle to anticipate when the AI provides correct and incorrect advice \cite{bansal2021does,schemmer2022should}. As the development of these assistance methods has so far been predominantly driven by an algorithmic perspective \cite{ehsan2021operationalizing}, necessary prerequisites from a human-centered perspective that contribute to enabling CTP have been underexplored. Consequently, researchers have recently started to argue for placing the human at the center of technology design \cite{ehsan2020human,liao2021human}. Thus, with our work, we aim to contribute to identifying essential elements in the interplay between humans and AI that have to be considered to enable effective decision-making.

\marginpar{%
\vspace{-165pt} 
\begin{minipage}{0.925\marginparwidth}
\centering
\resizebox{\marginparwidth}{!}{%

\includegraphics[]{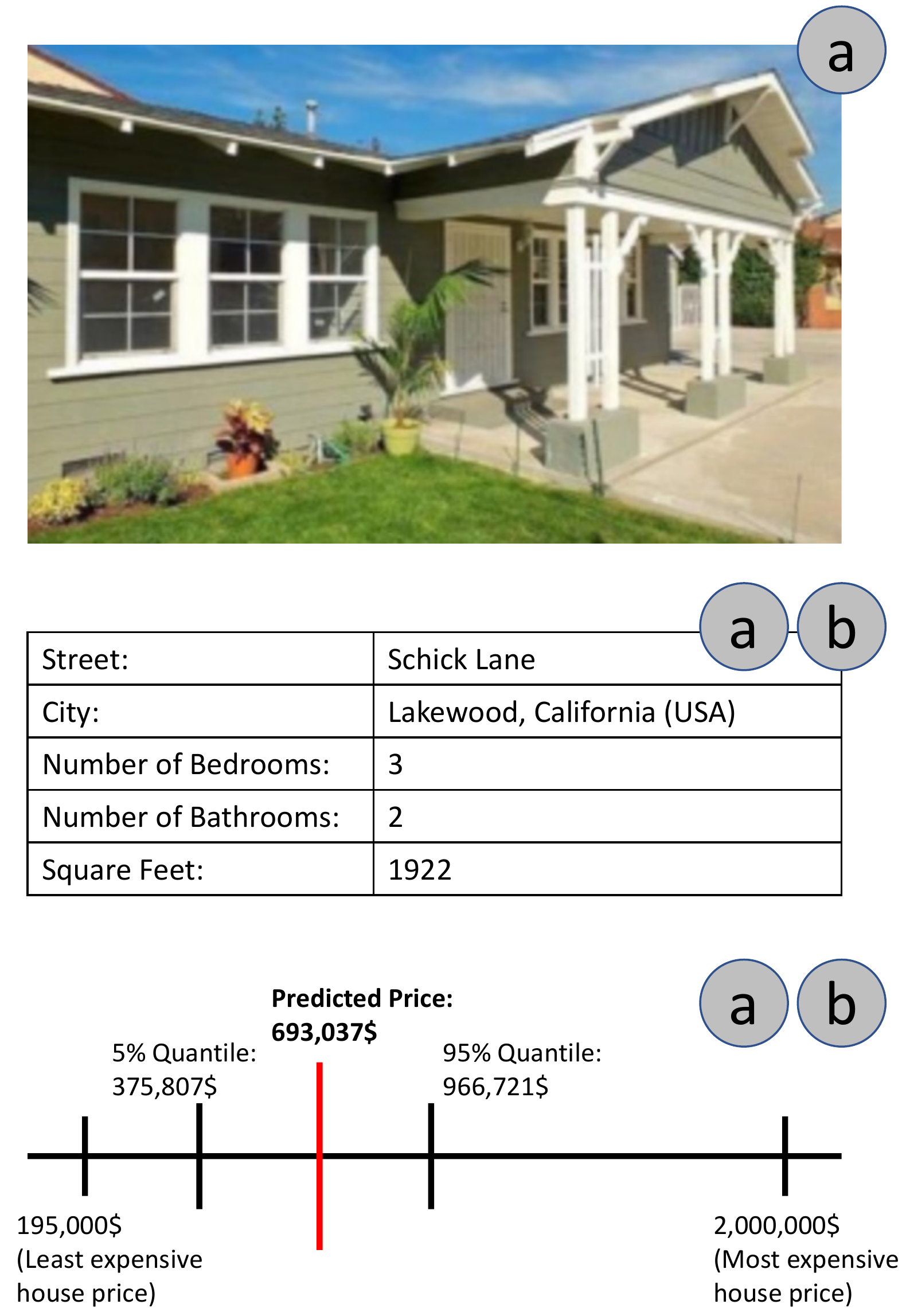}

}
\end{minipage}

{\vspace{0.5pc}\textbf{Figure 1:} An overview of the interface that contains the information provided to the participants in the online experiment. The \textit{UHCI treatment} provides all available information to the human (a). In the \textit{no UHCI treatment}, the image is withheld (b).}
\label{tab:study_overview}
}


\section{Methods}\label{chapter:methods}
We select a house prediction data set consisting of tabular data and house images \cite{ted8080_2019} for the online experiment. The data set consists of 15,474 instances, of which we allocate 80\% to the training and 20\% to the test set. Additionally, we draw a hold-out set of 15 properties from the test split as the samples for our experiment. We train an AI model---a random forest regression---only on the tabular features street, city, number of bedrooms, number of bathrooms, and square footage of the house. The image of the property is withheld from the AI model. It achieves a performance measured as the mean absolute error (MAE) of \$163,080 on the hold-out set, which is comparable to that on the entire test set. Additionally, based on the individual trees of the random forest, we generate a predictive distribution and display the 5\% and 95\% quantile as indicators for AI uncertainty.

The experiment consists of two treatments. In the first, participants are provided with information about each property's street, city, number of bedrooms, number of bathrooms, and square footage (\textit{no UHCI treatment}). In the second treatment, they are additionally provided with an image of the property (\textit{UHCI treatment}). With this image, they receive additional contextual information compared to the participants of the first treatment. Figure 1 displays the information provided in the respective treatments. We recruited participants via prolific.co and randomly assigned them to one of the two treatments. 

Before the actual set of tasks, participants in both treatments had to undergo an in-depth introduction to the data set and the task \cite{lai2020chicago,o1993judgemental}, including summary statistics about the properties' prices, followed by a question to verify their understanding. Additionally, we stressed that the AI did not have access to the image during training. 
After informing participants about the start of the actual decision-making task, the study procedure was as follows for each of the 15 instances: first, they were asked to provide a prediction on their own to prevent them from entering a state of low cognitive activation \cite{green2019principles}. Consequently, they received the AI's prediction together with its confidence estimate. Then, participants were asked to adjust the prediction of the AI in the best possible way. Finally, participants were asked to fill out a questionnaire to collect demographics after completing all instances. In general, participants received a base payment of 5 pounds with the incentive that the best 10\% would receive an additional pound. The whole task lasts approximately 30 minutes.

In total, we recruited 120 participants. To ensure the quality of the collected data, we removed participants entering house prices higher than the communicated maximum property price in the data set of \$2,000,000. Additionally, we identified outliers for removal using the median absolute deviation \cite{leys2013detecting,rousseeuw1993alternatives}. After applying these criteria, we collected the data from 101 participants over both conditions, of which 53 were in the \textit{no UHCI treatment} and 48 in the \textit{UHCI treatment}.

\section{Results And Discussion}

In \Cref{fig:performance_conditions}, we display the human and the AI-assisted performance for both conditions. We evaluate the significance of the results using the Student’s T-tests with Bonferroni correction. Its prerequisites have been verified ex-ante. Participants conducting the task alone in the \textit{no UHCI treatment} achieve a MAE of \$251,282, while the test persons in the \textit{UHCI treatment} yield a MAE of \$200,510. We observe a significant difference between both conditions without and with UHCI of \$50,772 $(t =4.6118 , p < 0.001)$. 

Looking at the team performance after adjusting the AI's prediction, we find that the human-AI team in the \textit{no UHCI treatment} achieves a MAE of \$160,095. In contrast, the human-AI team in the \textit{UHCI treatment} yields a performance in terms of MAE of \$148,009---a reduction of \$12,086. This performance improvement turns out to be significant on the 0.05 level $(t =2.9571 , p = 0.0155)$.
\setcounter{figure}{1}
\begin{figure}[h]
    \centering
    \includegraphics[width=\linewidth]{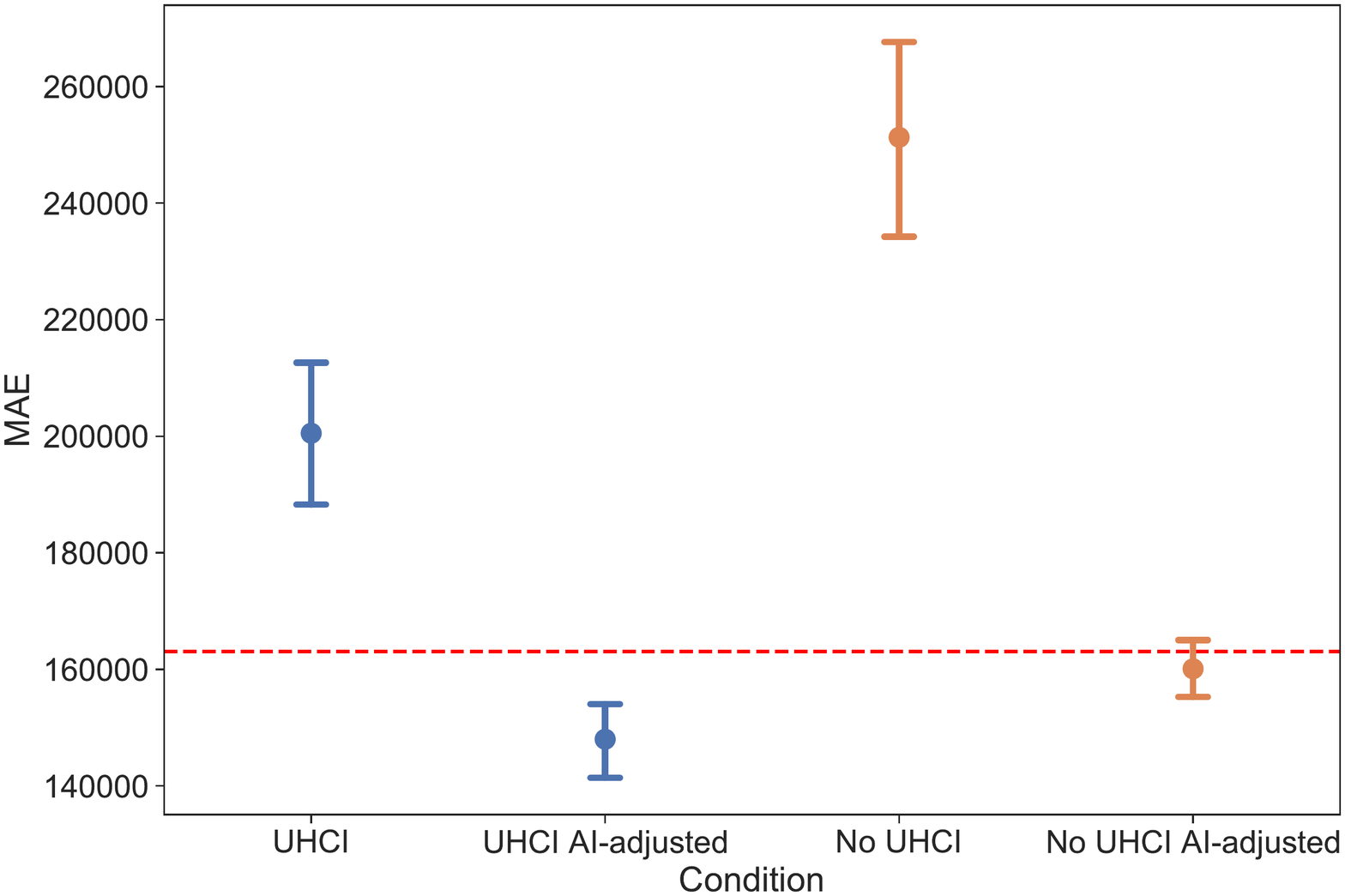}
    \caption{Performance results as MAE of the online experiment across conditions including 95\% confidence intervals. The red horizontal line denotes the AI performance.}
    \label{fig:performance_conditions}
\end{figure}
In both treatments, the human-AI teams outperform the AI alone (MAE: \$163,080). While the difference between the performance of the human-AI team in the \textit{UHCI treatment } is significant $(t =-4.6798 , p < 0.001)$, the difference in the \textit{no UHCI treatment} does not result in a significant improvement $(t =-1.1596 , p = 0.99)$. 
To summarize, we find that in the presence of UHCI, humans become capable of positively adjusting the AI predictions resulting in CTP. 

Regarding the general potential of human-AI teaming, our finding validates the results from \cite{bansal2021does,chu2020visual,lai2022human} by reaching CTP in an experimental study. Moreover, it has several implications for future research on human-AI decision-making in general and XAI-assisted decision-making in specific. For example, as humans tend to become more capable of correctly adjusting AI advice, they might also be able to better question additional information beyond sole confidence estimates, e.g., different explanations.

Future work should systematically identify additional sources of CP. From a human-centered perspective, not only information asymmetry but also skill differences could play a decisive role. Moreover, we hypothesize that CTP depends not only on CP but also on how well humans can utilize it in the decision-making process. Thus, human-centered design mechanisms to effectively combine AI and human decisions are needed to foster appropriate reliance on AI advice. Prior research on XAI has shown that a major challenge of XAI is the issue of over-trust \cite{bansal2021does,buccinca2021trust,schemmer2022influence}. Therefore, XAI needs to be designed taking a human-centered view to enable appropriate reliance and not solely increase trust. Additionally, future research needs to investigate the mutual effects between different XAI techniques and CP. In future work, we aim to formalize the notion of CP and conduct additional experiments to investigate the effect of XAI on appropriate reliance in the presence of CP.


\balance{} 

\bibliographystyle{SIGCHI-Reference-Format}
\bibliography{references}

\end{document}